\documentclass[twocolumn,english]{IEEEtran}
\usepackage[T1]{fontenc}
\usepackage{babel}
\usepackage{prettyref}
\usepackage{booktabs}
\usepackage{amsmath}
\usepackage{graphicx}
\usepackage[unicode=true,
 bookmarks=true,bookmarksnumbered=true,bookmarksopen=true,bookmarksopenlevel=1,
 breaklinks=false,pdfborder={0 0 0},backref=false,colorlinks=false]
 {hyperref}
\usepackage{breakurl}

\makeatletter

\providecommand{\tabularnewline}{\\}

 \let\oldforeign@language\foreign@language
 \DeclareRobustCommand{\foreign@language}[1]{%
   \lowercase{\oldforeign@language{#1}}}

\ifCLASSOPTIONcompsoc
\usepackage[caption=false,font=normalsize,labelfont=sf,textfont=sf]{subfig}
\else
\usepackage[caption=false,font=footnotesize]{subfig}
\fi

\usepackage{algpseudocode}

\makeatother

\begin{document}

\title{A GRASS GIS parallel module for radio-propagation predictions}

\author{Lucas~Benedi\v{c}i\v{c}, Felipe~A.~Cruz, Tsuyoshi~Hamada~and~Peter~Koro\v{s}ec%
\thanks{Lucas~Benedi\v{c}i\v{c} is with the Research and Development Department
of Telekom Slovenije, d.d., Ljubljana, Slovenia, e-mail: \protect\href{mailto:lucas.benedicic@telekom.si}{lucas.benedicic@telekom.si}.
Corresponding author.%
}%
\thanks{Felipe~A.~Cruz and Tsuyoshi~Hamada are with the Nagasaki Advanced
Computing Center, Nagasaki University, Nagasaki, Japan.%
}%
\thanks{Peter~Koro\v{s}ec is with the Computing Systems Department, Jo\v{z}ef
Stefan Institute, Ljubljana, Slovenia, e-mail: \protect\href{mailto:peter.korosec@ijs.si}{peter.korosec@ijs.si}.

\emph{Preprint submitted to Taylor \& Francis.}%
}}

\markboth{arXiv.org preprint}{Benedi\v{c}i\v{c} \MakeLowercase{\emph{et al.}}:
A GRASS GIS parallel module for radio-propagation predictions}
\maketitle
\begin{abstract}
Geographical information systems are ideal candidates for the application
of parallel programming techniques, mainly because they usually handle
large data sets. To help us deal with complex calculations over such
data sets, we investigated the performance constraints of a classic
master-worker parallel paradigm over a message-passing communication
model. To this end, we present a new approach that employs an external
database in order to improve the calculation/communication overlap,
thus reducing the idle times for the worker processes. The presented
approach is implemented as part of a parallel radio-coverage prediction
tool for the GRASS environment. The prediction calculation employs
digital elevation models and land-usage data in order to analyze the
radio coverage of a geographical area. We provide an extended analysis
of the experimental results, which are based on real data from an
LTE network currently deployed in Slovenia. Based on the results of
the experiments, which were performed on a computer cluster, the new
approach exhibits better scalability than the traditional master-worker
approach. We successfully tackled real-world data sets, while
greatly reducing the processing time and saturating the hardware utilization.\end{abstract}

\section{Introduction}

\IEEEPARstart{A}{lthough} Gordon Moore's well-known and often cited
prediction still holds \cite{Moore_Cramming_more_components_onto_integrated_circuits:1998},
the fact is that for the past few years, CPU speeds have hardly been
improving. Instead, the number of cores within a single CPU is increasing.
This situation poses a challenge for software development in general
and research in particular: a hardware upgrade will, most of the time,
fail to double the serial execution speed of its predecessor. However,
since this commodity hardware is present in practically all modern
desktop computers, it creates an opportunity for the parallel exploitation
of these computing resources to enhance the performance of complex
algorithms over large data sets. The challenge is thus to deliver
the computing power of multi-core systems in order to tackle a computationally
time-consuming problem, the completion of which is unfeasible using
traditional serial approaches. Moreover, by accessing many such computing
nodes through a network connection, even more possibilities are available.

A traditional approach when dealing with computationally expensive
problem solving is to simplify the models in order to be able to execute
their calculations within a feasible amount of time. Clearly, this
method increases the introduced error level, which is not an option
for a certain group of simulations, e.g., those dealing with disaster
contingency planning and decision support \cite{Huang_Using_adaptively_coupled_models_and_high_performance_computing_for_enabling_the_computability_of_dust_storm_forecasting:2012,Yin_A_framework_for_integrating_GIS_and_parallel_computing_for_spatial_control_problems_a_case_study_of_wildfire_dontrol:2012}.
The conducted simulations during the planning phase of a radio network
also belong to this group. Their results are the basis for the decision
making prior to physically installing the base stations and antennas
that will cover a certain geographical area. A greater deviation of
these results increases the probability of making the wrong decisions
at the time of the installation, which may considerably increase the
costs or even cause mobile-network operators to incur losses.

Various groups have successfully deployed high-performance computing
(HPC) systems and techniques to solve different problems dealing with
spatial data \cite{Akhter_Porting_GRASS_raster_module_to_distributed_computing:2007,Armstrong_Using_a_computational_grid_for_geographic_information_analysis:2005,Guan_A_parallel_computing_approach_to_fast_geostatistical_areal_interpolation:2011,Huang_Using_adaptively_coupled_models_and_high_performance_computing_for_enabling_the_computability_of_dust_storm_forecasting:2012,Li_Parallel_cellular_automata_for_large_scale_urban_simulation_using_load_balancing_techniques:2010,Osterman_CUDA_on_GRASS:2012,Tabik-High_performance_three_horizon_composition_algorithm_for_large_scale_terrains:2011,Tabik-Optimal_tilt_and_orientation_maps_a_multi_algorithm_approach_for_heterogeneous_multicore_GPU_systems:2013,Tabik_Simultaneous_computation_of_total_viewshed_on_large_high_resolution_grids:2012,Widener_Developing_a_parallel_computational_implementation_of_AMOEBA:2012,Yin_A_framework_for_integrating_GIS_and_parallel_computing_for_spatial_control_problems_a_case_study_of_wildfire_dontrol:2012}.
This research has confirmed that a parallel paradigm such as master-worker,
techniques like work pool (or task farming) and spatial-block partitioning
are applicable when dealing with parallel implementations over large
spatial data sets. However, it is well known that parallel programming
and HPC often call for area experts in order to integrate these practices
into a given environment \cite{Clematis_High_performance_computing_with_geographical_data:2003}.
Moreover, the wide range of options currently available creates even
more barriers for general users wanting to benefit from HPC.

In this paper, we combine some of the known principles of HPC and
introduce a new approach in order to improve the performance speed
of a GIS module for radio-propagation predictions. The efficiency
improvement is based on overlapping process execution and communication
in order to minimize the idle time of the worker processes and thus
improve the overall efficiency of the system. To this end, we save
the intermediate calculation results into an external database (DB)
instead of sending them back to the master process. We implement this
approach as part of a parallel radio-prediction tool (PRATO) for the
open-source Geographic Resources Analysis Support System (GRASS) \cite{Neteler_Open_source_GIS_a_GRASS_GIS_approach}.
For its architecture, we have focused on scalability, clean design
and the openness of the tool, inspired by the GRASS GIS. This makes
it an ideal candidate for demonstrating the benefits and drawbacks
of several reviewed patterns, while tackling the radio-coverage predictions
of big problem instances, e.g., real mobile networks containing thousands
of transmitters over high-resolution terrains, and big-scale simulations
covering the whole country.

\subsection{Objectives\label{sub:Objectives}}

In order to assess the benefits and drawbacks of various reviewed
approaches from the performance point of view, we evaluate PRATO in
a distributed computing environment. Furthermore, by presenting a
detailed description of its design and implementation, we provide
an analysis of the patterns achieving higher efficiency levels, so
that they can be adopted for general task parallelization in the GRASS
GIS.

The paper is organized as follows. Section \ref{sec:Related-work}
gives an overview of the relevant publications, describing how they
relate to our work. Section \ref{sec:Description-of-the-radio-coverage-prediction-tool}
gives a description of the radio-coverage prediction problem, including
the radio-propagation model. Section \ref{sec:Design-and-implementation}
concentrates on the design principles and implementation details of
the radio-propagation tool, for the serial and parallel versions.
Section \ref{sec:Simulations} discusses the experimental results
and their analysis. Finally, Section \ref{sec:Conclusion} draws some
conclusions.

\section{Related work \label{sec:Related-work}}

The task-parallelization problem within the GRASS environment has
been addressed by several authors in a variety of studies. For example,
in \cite{Campos_Parallel_modelling_in_GIS:2012}, the authors present
a collection of GRASS modules for a watershed analysis. Their work
concentrates on different ways of slicing raster maps to take advantage
of a Message Passing Interface (MPI) implementation.

In the field of high-performance computing, the authors of~\cite{Akhter_Porting_GRASS_raster_module_to_distributed_computing:2007}
presented implementation examples of a GRASS raster module, used to
process vegetation indexes for satellite images, for MPI and Ninf-G
environments. The authors acknowledge a limitation in the performance
of their MPI implementation for big processing jobs. The restriction
appears due to the computing nodes being fixed to a specific spatial
range, since the input data are equally distributed among worker processes,
creating an obstacle for load balancing in heterogeneous environments.

Using a master-worker technique, the work by \cite{Huang-Explorations_of_the_implementation_of_a_parallel_IDW_algorithm_in_a_Linux_cluster:2011}
abstracts the GRASS data types into its own \emph{struct} and MPI
data types, thus not requiring the GRASS in the worker nodes. The
data are evenly distributed by row among the workers, with each one
receiving an exclusive column extent to work on. The test cluster
contains heterogeneous hardware configurations. The authors note that
data-set size is bounded by the amount of memory on each of the nodes,
since they allocate the memory for the whole map as part of the set-up
stage, before starting the calculation. Regarding the data sets during
the simulations, the largest one contains 3,265,110 points. They conclude
that the data-set size should be large enough for the communication
overhead to be hidden by the calculation time, so that the parallelization
pays off.

In \cite{Tabik-High_performance_three_horizon_composition_algorithm_for_large_scale_terrains:2011},
the authors employ a master-worker approach, using one worker process
per worker node. The complete exploitation of the computing resources
of a single computing node is achieved with OpenMP. The experimental
environment features one host. The horizon-composition algorithm presents
no calculation dependency among the spatial blocks. Consequently,
the digital elevation model (DEM) may be divided into separate blocks
to be independently calculated by each worker process. The authors
present an improved algorithm that can also be used to accelerate
other applications like visibility maps. The tasks are dynamically
assigned to idle processes using a task-farming paradigm over the
MPI. 

Also, in \cite{Tabik-Optimal_tilt_and_orientation_maps_a_multi_algorithm_approach_for_heterogeneous_multicore_GPU_systems:2013}
there is no calculation dependency among the spatial blocks. The experimental
evaluation is made over multiple cores of one CPU and a GPU, communicated
using a master-worker setup.

In \cite{Yin_A_framework_for_integrating_GIS_and_parallel_computing_for_spatial_control_problems_a_case_study_of_wildfire_dontrol:2012},
the authors present a parallel framework for GIS integration. Based
on the principle of spatial dependency, they lower the calculation
processing time by backing it with a knowledge database, delivering
the heavy calculation load to the parallel back-end if a specific
problem instance is not found in the database. There is an additional
effort to achieve the presented goals, since the implementation of
a fully functional GIS (or ``thick GIS'' as the authors call it)
is required on both the desktop client and in the parallel environment.

An agent-based approach for simulating spatial interactions is presented
in \cite{Gong_Parallel_agent_based_simulation_of_individual_level_spatial_interactions_within_a_multicore_computing_environment:2012}.
The authors' approach decomposes the entire landscape into equally-sized
regions, i.e., a spatial-block division as in \cite{Tabik-High_performance_three_horizon_composition_algorithm_for_large_scale_terrains:2011},
which are in turn processed by a different core of a multi-core CPU.
This work uses multi-core CPUs instead of a computing cluster.

Some years ago, grid computing received the attention of the research
community as a way of accessing the extra computational power needed
for the spatial analysis of large data sets \cite{Armstrong_Using_a_computational_grid_for_geographic_information_analysis:2005,Vouk_Cloud_computing_issues_research_and_implementations:2008,Wang_A_cybergis_framework_for_the_synthesis_of_cyberinfrastructure_GIS_and_spatial_analysis:2010}.
However, several obstacles are still preventing this technology from
being more widely used. Namely, its adoption requires not only hardware
and software compromises with respect to the involved parts, but also
a behavioral change at the human level \cite{Armstrong_Using_a_computational_grid_for_geographic_information_analysis:2005}.

\section{Radio-coverage prediction for mobile networks \label{sec:Description-of-the-radio-coverage-prediction-tool}}

\subsection{Background}

The coverage planning of radio networks is a key problem that all
mobile operators have to deal with. Moreover, it has proven to be
a fundamental issue, not only in LTE networks, but also in other standards
for mobile communications \cite{Saleh_On_the_coveraga_extension_in_LTE_networks:2010,Shabbir_Comparison_of_radio_propagation_models:2011,Siomina:Minimum.pilot.power.for.service.coverage,Valcarce_Applying.FDTD.to.the.coverage.prediction.of.WiMAX:2009}.
One of the primary objectives of mobile-network planning is to efficiently
use the allocated frequency band to ensure that some geographical
area of interest can be satisfactorily reached with the base stations
of the network. To this end, radio-coverage prediction tools are of
great importance as they allow network engineers to test different
network configurations before physically implementing the changes.
Nevertheless, radio-coverage prediction is a complex task, mainly
due to the several combinations of hardware and configuration parameters
that have to be analyzed in the context of different environments.
The complexity of the problem means that radio-coverage prediction
is a computationally-intensive and time-consuming task, hence the
importance of using fast and accurate tools (see Section \ref{sub:Computational-complexity}
for a complexity analysis of the algorithm). Additionally, since the
number of deployed transmitters keeps growing with the adoption of
modern standards \cite{Saleh_On_the_coveraga_extension_in_LTE_networks:2010},
there is a clear need for a radio-propagation tool that is able to
cope with larger work loads in a feasible amount of time (see Section
\ref{sub:Computational-complexity} for the running time of the serial
version).

In this work, we present PRATO: a high-performance radio-propagation
prediction tool for GSM (2G), UMTS (3G) and LTE (4G) radio networks.
It is implemented as a module of the GRASS GIS. It can be used for
planning the different phases of a new radio-network installation,
as well as a support tool for maintenance activities related to network
troubleshooting in general and optimization in particular. Specifically,
automatic radio-coverage optimization requires the evaluation of millions
of radio-propagation predictions in order to find a good solution
set, which is unfeasible using other serial implementations of academic
or commercial tools \cite{Ozimek_Open.source.radio.coverage.prediction:2010,Mehlfuhrer_The_Vienna_LTE_Simulators_enabling_reproducibility_in_wireless_communications_research:2011,Piro_Simulating_LTE_cellular_systems_an_open_source_framework:2011}.

As a reference implementation, we used the publicly available radio-coverage
prediction tool, developed in \cite{Ozimek_Open.source.radio.coverage.prediction:2010}.
The authors of this work developed a modular radio-coverage tool that
performs separate calculations for radio-signal path loss and antenna
radiation patterns, also taking into account different configuration
parameters, such as antenna tilting, azimuth and height. The output
result, saved as a raster map, is the maximum signal level over the
target area, in which each point represents the received signal from
the best serving transmitter. This work implements some well-known
radio-propagation models, e.g., Okumura-Hata~\cite{Hata_Empirical_formula_for_propagation_loss_in_land_mobile_radio_services:1980}
and COST 231~\cite{Cichon_Propagation.prediction.models:1995}. The
latter is explained in more detail in Section \ref{sub:COST-231-model}.
Regarding the accuracy of the predicted values, the authors~\cite{Ozimek_Open.source.radio.coverage.prediction:2010}
report comparable results to those of a state-of-the-art commercial
tool. To ensure that our implementation is completely compliant with
the previously mentioned reference, we have designed a comparison
test that consists of running both tools with the same set of input
parameters. The test results from PRATO and the reference implementation
were identical in all the tested cases.

\subsection{Propagation modeling\label{sub:COST-231-model}}

PRATO uses the COST-231 Walfisch-Ikegami radio-propagation model \cite{Shabbir_Comparison_of_radio_propagation_models:2011},
which was introduced as an extension of the well-known COST Hata model
\cite{Sarkar_Survey_of_radio_propagation_models:2003}. The suitability
of this model comes from the fact that it distinguishes between line-of-sight
(LOS) and non-line-of-sight (NLOS) conditions.

In this work, as well as in the reference implementation~\cite{Ozimek_Open.source.radio.coverage.prediction:2010},
the terrain profile is used for the LOS determination. In this context,
a NLOS situation appears when the first Fresnel zone is obscured by
at least one obstacle \cite{Xia-Radio_propagation_characteristics_for_line_of_sight_microcellular_and_personal_communications:1993}.
We include a correction factor, based on the land usage (clutter data),
for accurately predicting the signal-loss effects due to foliage,
buildings and other fabricated structures. This technique is also
adopted by other propagation models for radio networks, like the artificial
neural networks macro-cell model developed in~\cite{Neskovic_Microcell_electric_field_strength_prediction_model:2010}.
Consequently, we introduce an extra term for signal loss due to clutter
($L_{\textrm{CLUT}}$) to the Walfisch-Ikegami model~\cite{Shabbir_Comparison_of_radio_propagation_models:2011},
defining the path loss as

\begin{equation}
PL(d)=L_{0}(d)+L_{\textrm{CLUT}}+\begin{cases}
PL_{\mathrm{LOS}}(d)\\
PL_{\textrm{NLOS}}(d)
\end{cases},\label{eq:cost231}
\end{equation}

\noindent where $L_{0}(d)$ is the attenuation in free space and is
defined as

\begin{equation}
L_{0}(d)=32.45+20\log(d)+20log(F).\label{eq:cost231-L0}
\end{equation}

\noindent If there is LOS between the transmitter antenna and the
mobile, the path loss $PL_{\mathrm{LOS}}(d)$ is defined as

\begin{equation}
PL_{\textrm{LOS}}(d)=42.64+26\log(d)+20\log(F),\label{eq:cost231_LOS}
\end{equation}
whereas the path loss for NLOS conditions is determined as

\begin{equation}
PL_{\textrm{NLOS}}(d)=L_{\textrm{RTS}}+L_{\textrm{MSD}}.\label{eq:cost231_NLOS}
\end{equation}
Here, $d$ is the distance (in kilometers) from the transmitter to
the receiver point, $F$ is the frequency (in MHz), $L_{\textrm{RTS}}$
represents the diffraction from rooftop to the street, and $L_{\textrm{MSD}}$
represents the diffraction loss due to multiple obstacles. Consequently,
the total path loss from the antenna to the mobile device is calculated
as in Equation~(\ref{eq:cost231}), where the attenuation in free
space and due to clutter are also taken into account.

\section{Design and implementation \label{sec:Design-and-implementation}}

\subsection{Design of the serial version}

\begin{figure}[tbh]
\centering

\includegraphics[width=1\columnwidth]{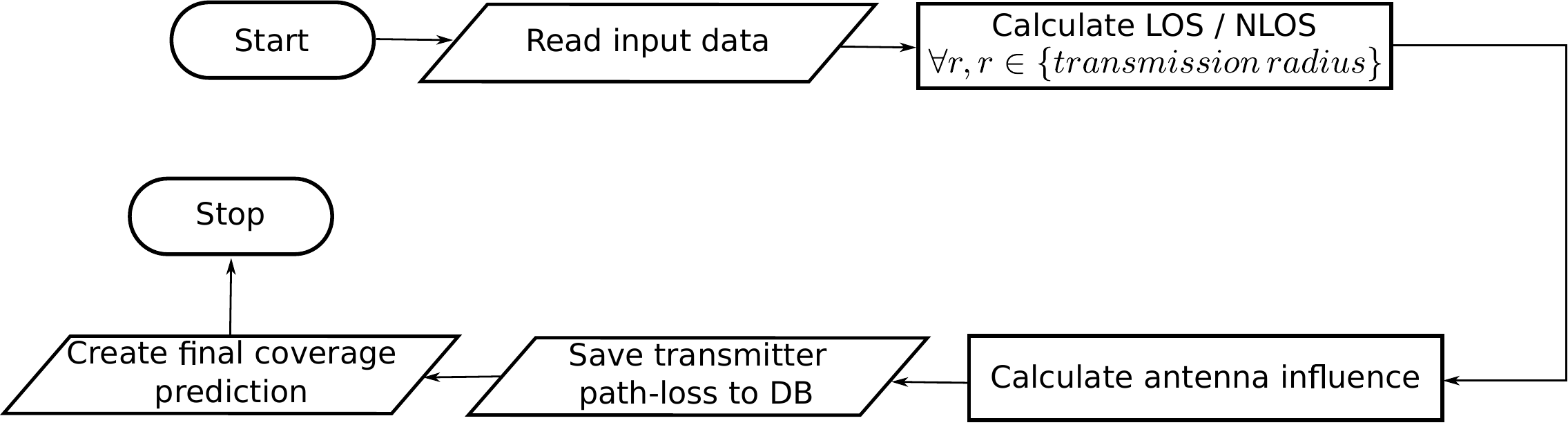}

\caption{Flow diagram of the serial version. \label{fig:serial_version_flow_diagram}}
\end{figure}

This section describes the different functions contained in the serial
version of PRATO, which is implemented as a GRASS module. Their connections
and data flow are depicted in \prettyref{fig:serial_version_flow_diagram},
where the parallelograms of the flow diagram represent the input/output
(I/O) operations. 

Our design follows a similar internal organization as the radio-planning
tool presented in \cite{Ozimek_Open.source.radio.coverage.prediction:2010},
but with some important differences. First, the modular design was
avoided in order to prevent the overhead of I/O operations between
the components of a modular architecture. Second, our approach employs
a direct connection to an external database server for intermediate
result saving, instead of the slow, built-in GRASS database drivers.
To explicitly avoid tight coupling with a specific database vendor,
the generated output is formatted in plain text, which is then forwarded
to the DB. Any further processing is achieved by issuing a query over
the database tables that contain the partial results for each of the
processed transmitters.

\begin{figure*}[tbh]
\begin{minipage}[t]{0.49\textwidth}%
\centering

\includegraphics[width=0.8\columnwidth]{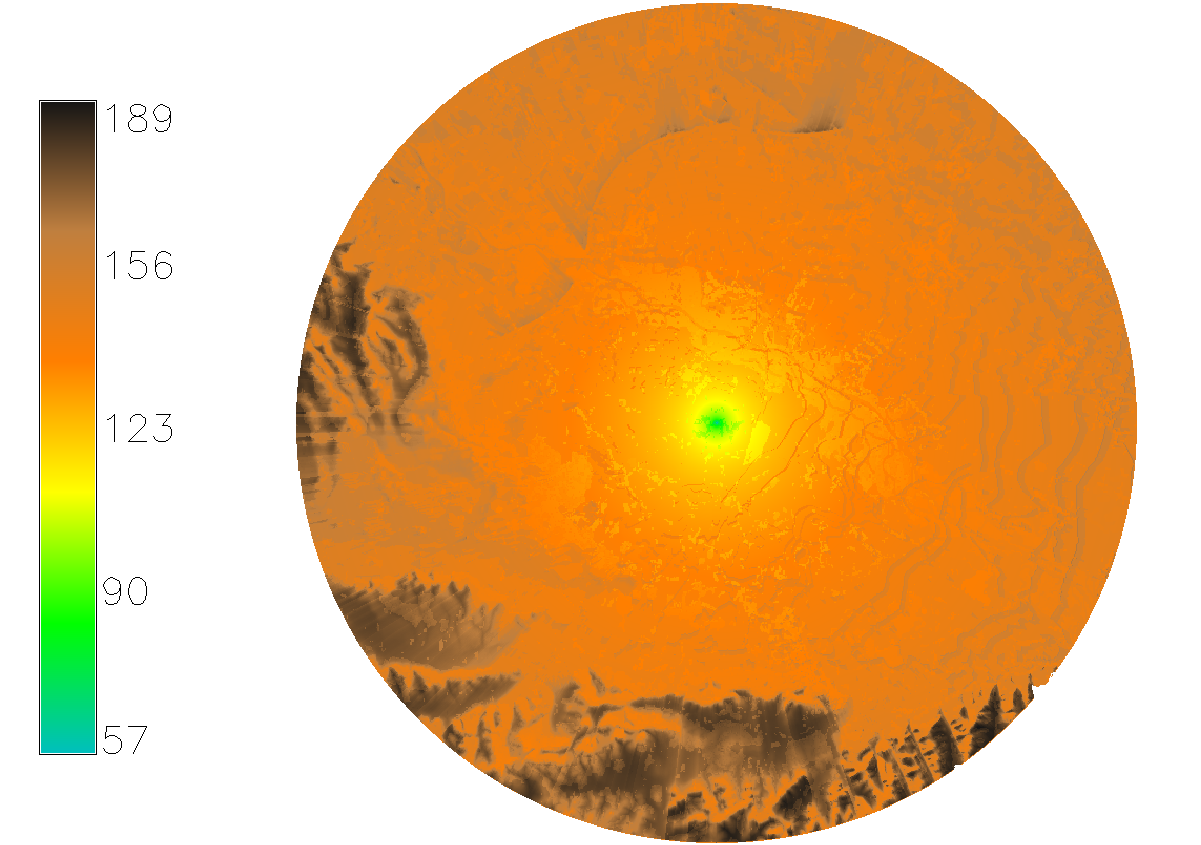}

\caption{Example of raster map, showing the result of a path-loss
calculation from an isotropic source.\label{fig:path_loss-example}}
\end{minipage}\hfill{}%
\begin{minipage}[t]{0.49\textwidth}%
\centering

\includegraphics[width=0.8\columnwidth]{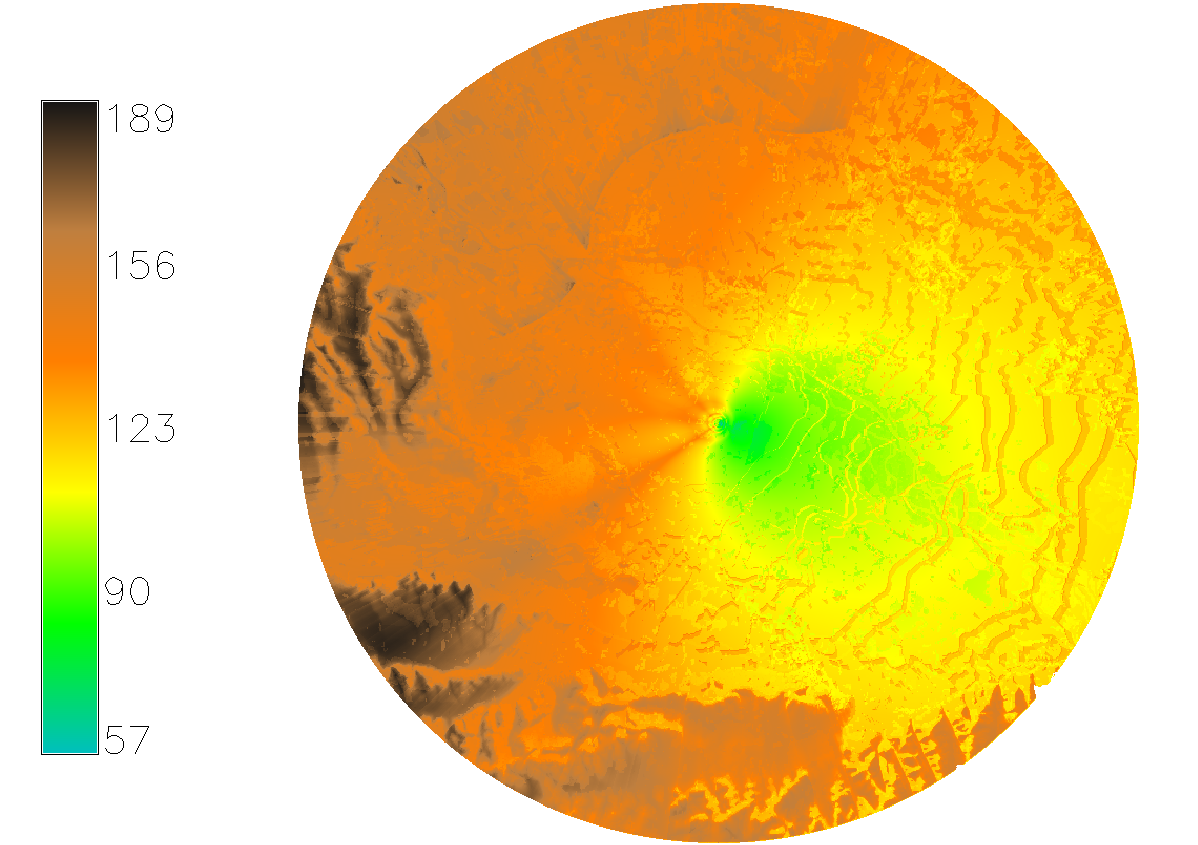}

\caption{Example of raster map, showing the antenna influence
over the isotropic path-loss result, as depicted in \prettyref{fig:path_loss-example}.\label{fig:antenna-example}}
\end{minipage}
\end{figure*}

\subsubsection{Isotropic path-loss calculation\label{sub:Path-loss-for-isotrophic-source}}

This step starts by calculating which receiver points, $r$, are within
the specified transmission radius (see ``\emph{transmission radius}''
in \prettyref{fig:serial_version_flow_diagram}). The transmission
radius is defined around each transmitter in order to limit the radio-propagation
calculation to a reasonable distance. For these points, the LOS and
NLOS conditions are calculated with respect to the transmitter (see
``Calculate LOS/NLOS'' in \prettyref{fig:serial_version_flow_diagram}).
The following step consists of calculating the path loss for an isotropic
source (or omni antenna). This calculation is performed by applying
the Walfisch-Ikegami model, which was previously defined in Equation
(\ref{eq:cost231}), to each of the points within the transmission
radius around the transmitter.

\prettyref{fig:path_loss-example} shows an example result of the
isotropic path-loss calculation, only including the map area within
the transmission radius. The color scale is given in dB, indicating
the signal loss from the isotropic source of the transmitter, located
at the center. Notice the hilly terrain is clearly distinguished due
to LOS and NLOS conditions from the signal source.

\subsubsection{Antenna diagram influence\label{sub:Antenna-diagram-influence}}

This step considers the antenna radiation diagram of the current transmitter
and its influence over the isotropic path-loss calculation (see ``Calculate
antenna influence'' in \prettyref{fig:serial_version_flow_diagram}).
Working on the in-memory results generated by the previous step, the
radiation diagram of the antenna is taken into account, including
the beam direction, the electrical and the mechanical tilt. \prettyref{fig:antenna-example}
shows the map area within the transmission radius, where this calculation
step was applied to the results from \prettyref{fig:path_loss-example}.
Notice the distortion of the signal propagation that the antenna has
introduced.

\subsubsection{Transmitter path-loss prediction\label{sub:Transmitter-path-loss-prediction}}

In this step, the path-loss prediction of the transmitter is saved
in its own database table (see ``Save transmitter path-loss to DB''
in \prettyref{fig:serial_version_flow_diagram}). This is accomplished
by connecting the standard output of the developed module with the
standard input of a database client. Naturally, the generated plain
text should be understood by the DB itself.

\subsubsection{Coverage prediction\label{sub:Final-coverage-prediction}}

\begin{figure}[tbh]
\centering

\includegraphics[width=0.7\columnwidth]{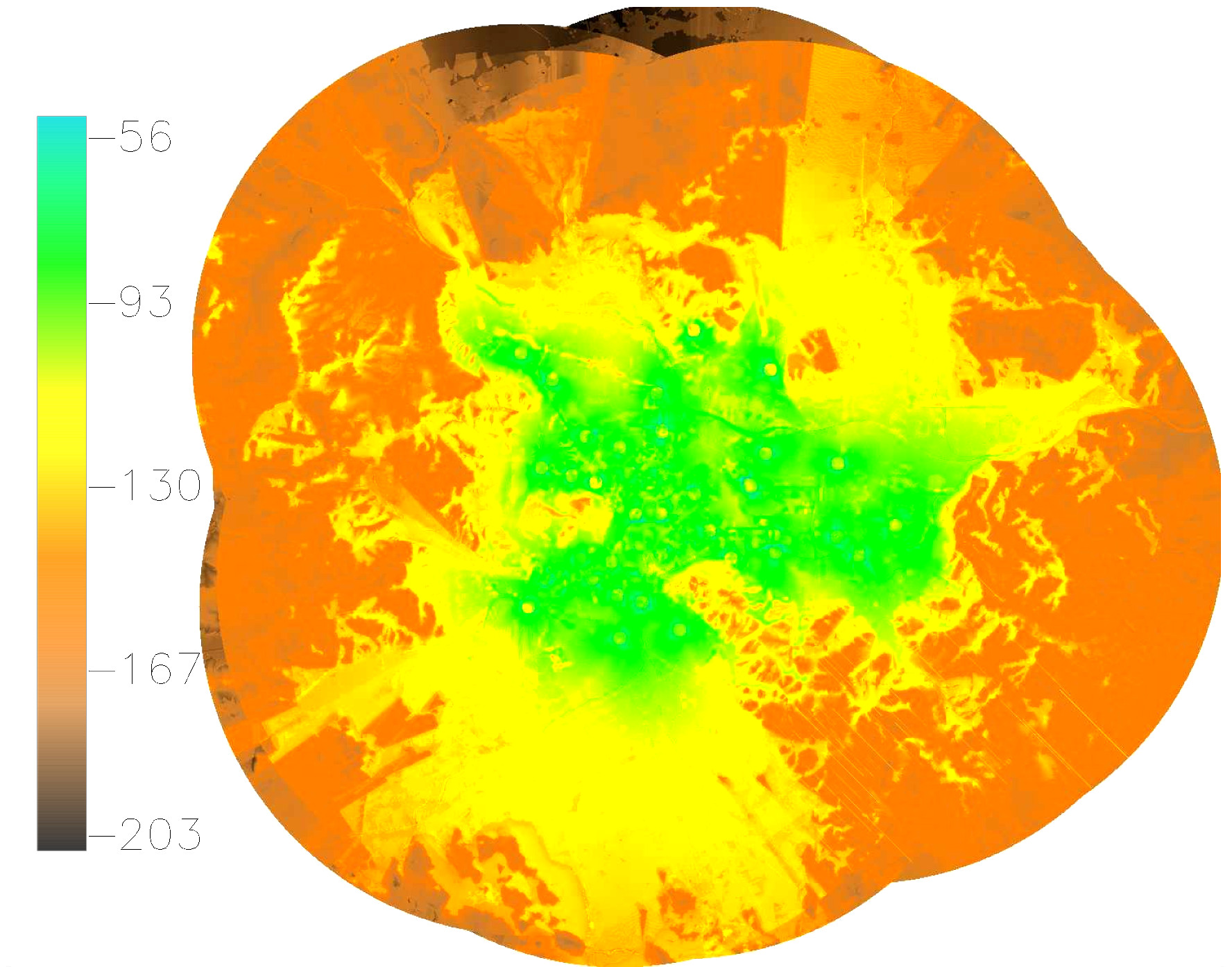}

\caption{Example of a raster map, displaying the final coverage
prediction of 136 transmitters over a geographical area. The color
scale is given in dBm, indicating the received signal strength. Darker
colors denote areas with a reduced signal due to the fading effect
of the hilly terrain and clutter. \label{fig:output_raster_example}}
\end{figure}

The final radio-coverage prediction, containing the aggregation of
the partial path-loss results of the involved transmitters, is created
in this step (see ``Create final coverage prediction'' in \prettyref{fig:serial_version_flow_diagram}).
The received signal strength from each of the transmitters is calculated
as the difference between its transmit power and the path loss for
the receiver's corresponding position. This is done by executing an
SQL query over the tables containing the path-loss predictions of
each of the processed transmitters. Finally, the output is generated,
using the GRASS built-in modules $v.in.ascii$ and $v.to.rast$, which
create a raster map using the query results as the input. The final
raster map contains the maximum received signal strength for each
individual point, as shown in \prettyref{fig:output_raster_example}.
In this case, the color scale is given in dBm, indicating the strongest
received signal strength from the transmitters.

\subsection{Computational complexity of the radio-coverage algorithm \label{sub:Computational-complexity}}

\begin{table}[tbh]
\centering

\caption{Pseudo code of the radio-coverage prediction algorithm. The time complexity
is given per line.\label{tab:Pseudocode-radio_coverage_algorithm}}

\begin{algorithmic}
\State $DEM \gets$ Digital Elevation Model (DEM) of the whole area.
\Comment $O(M)$
\State $Clutter \gets$ signal Losses due to land usage of the whole area.
\Comment $O(M)$
\State $T \gets$ transmitter configuration data.
\Comment $O(n)$
\ForAll{$t \in T$}
	\Comment $O(n \cdot m^2)$
	\State $DEM_t \gets $ DEM area within transmission radius of $t$
	\Comment $O(m)$
	\State $Clut_t \gets $ Clutter area within transmission radius $t$
	\Comment $O(m)$
	\State $LoS_t \gets$ LineOfSight ($DEM_t$)
	\Comment $O(m^2)$
	\State $PL_t \gets$ PathLoss ($DEM_t, Clut_t, LoS_t$)
	\Comment $O(m^2)$
	\State $Diag_t \gets $ Antenna diagram of $t$ 
	\Comment $O(1)$
	\State $PL_t \gets$ AntennaInfluence ($Diag_t, PL_t$)
	\Comment $O(m)$
\EndFor
\ForAll{$t \in T$}
	\Comment $O(n \cdot m)$
	\State $CoveragePrediction \gets$ PathLossAggregation ($t, PL_t$)
	\Comment $O(m)$
\EndFor
\State \Return $CoveragePrediction$
\end{algorithmic}
\end{table}

In this section, the time complexity of the radio-coverage prediction
algorithm is presented, for which the pseudo code is listed
in Table \ref{tab:Pseudocode-radio_coverage_algorithm}.

The algorithm starts by loading the input, i.e., the DEM and the clutter
data. Both regular square grids (RSGs) should account for the same
area and resolution, consequently containing the same number of pixels,
$M$. The transmitter data is then loaded into set $T$, the cardinality
of which is denoted as $n=|T|$. For each transmitter $t\in T$, a
smaller subarea of the DEM and clutter data (denoted $DEM_{t}$ and
$Clut_{t}$, respectively) is delimited around $t$, based on a given
transmission radius. The number of pixels within this sub-area is
denoted as $m$, and its value is the same for all $t\in T$. The
visibility for an RSG cell is computed using the \emph{LineOfSight}
function, by walking from the antenna of the transmitter to the given
element, along the elements intersected by a LOS, until either the
visibility is blocked, or the target is reached~\cite{DeFloriani-Applications_of_computational_geometry_to_geographic_information_systems:1999}.
Regarding the \emph{PathLoss }function, whenever a receiver point
is in NLOS, the walking path from the transmitter has to be inspected
for obstacles, calculating the diffraction losses for each of them,
i.e., $L_{\textrm{MSD}}$ from Equation (\ref{eq:cost231_NLOS}).
Hence, its quadratic complexity, which dominates the complexity of
the algorithm, together with \emph{LineOfSight}, resulting in an algorithmic
complexity denoted by

\begin{equation}
O(M+n\cdot m^{2}).
\end{equation}

Although $n$ will generally be many orders of magnitude smaller than
$m^{2}$, its computational-time complexity is relevant for practical
use. For example, assuming the radio-coverage prediction for one transmitter
completes in around 15~seconds using a serial implementation, the
prediction for a mobile network comprising 10,240 transmitters would
have an execution time of almost two days.

\subsection{Multi-paradigm parallel programming}

The implementation methodology adopted for PRATO follows a multi-paradigm,
parallel programming approach in order to fully exploit the resources
of each of the nodes in a computing cluster. This approach combines
a master-worker paradigm with an external DB. To efficiently use a
shared memory multi-processor on the worker side, and to effectively
overlap the calculation and communication, PRATO uses POSIX threads~\cite{Butenhof_Programming.with.POSIX.threads:1997}.

To use the computing resources of a distributed memory system, such
as a cluster of processors, PRATO uses the MPI~\cite{Gropp_Using_MPI:1999}.
The MPI is a message-passing standard that defines the syntax and
semantics designed to function on a wide variety of parallel computers.
The MPI enables multiple processes, running on different processors
of a computer cluster, to communicate with each other. It was designed
for high performance on both massively parallel machines and on workstation
clusters.

In order to make the text clearer and to differentiate between the
programming paradigms used from here on, we will refer to a POSIX
thread simply as a `thread' and a MPI process as a `process'.

\subsection{Design of the parallel version\label{sub:Design-parallel}}

By maintaining our focus on the practical usability and performance
of PRATO, we are introducing a parallel implementation to overcome
the computational-time constraints that prevent a serial implementation
of the radio-coverage prediction algorithm from tackling big problem
instances in a feasible amount of time.

A major drawback of the GRASS as a parallelization environment is
that it is not thread-safe, meaning that concurrent changes to the
same data set have an undefined behavior~\cite{Blazek_GRASS_server:2004}.
One technique to overcome this problem is to abstract the spatial
data from the GRASS. For example, in~\cite{Huang-Explorations_of_the_implementation_of_a_parallel_IDW_algorithm_in_a_Linux_cluster:2011},
the authors achieve the GRASS abstraction by introducing a \emph{Point
}structure with four \emph{double} attributes, where each pixel of
the RSG is mapped to an instance of this structure. Another possibility
is for one of the processes, e.g., the master, to read entire rows
or columns of data before dispatching them for processing to the workers~\cite{Akhter_Porting_GRASS_raster_module_to_distributed_computing:2007,Huang-Explorations_of_the_implementation_of_a_parallel_IDW_algorithm_in_a_Linux_cluster:2011}.
In this case, an independence between row/column calculations is required,
which is a problem-specific property. In our case, we propose to achieve
the GRASS abstraction by loading the spatial data into a 2D matrix
(or matrices) of basic data-type elements, e.g., \emph{float} or \emph{double}
depending on the desired accuracy. The geographical location of each
element is calculated as the geographical location of the matrix plus
the element offset within it. The advantage of this technique is having
the geographical location of each pixel readily available with a minimum
memory footprint. Moreover, a convenient consequence of this abstraction
schema is that worker processes are completely independent of the
GRASS, thus significantly simplifying the deployment of the parallel
implementation over multiple computing hosts.

In the area of geographical information science, the master-worker
paradigm has been successfully applied by several authors \cite{Akhter-GRASS_GIS_on_high_performance_computing_with_MPI_OpenMP_and_Ninf-G:2010,Akhter_Porting_GRASS_raster_module_to_distributed_computing:2007,Campos_Parallel_modelling_in_GIS:2012,Guan_A_parallel_computing_approach_to_fast_geostatistical_areal_interpolation:2011,Huang-Explorations_of_the_implementation_of_a_parallel_IDW_algorithm_in_a_Linux_cluster:2011,Tabik-High_performance_three_horizon_composition_algorithm_for_large_scale_terrains:2011,Tabik-Optimal_tilt_and_orientation_maps_a_multi_algorithm_approach_for_heterogeneous_multicore_GPU_systems:2013}.
However, sometimes this technique presents certain issues that prevent
the full exploitation of the available computing resources when deployed
over several networked computers. Additionally, such issues are difficult
to measure when the parallelization involves only one computing node
\cite{Tabik-High_performance_three_horizon_composition_algorithm_for_large_scale_terrains:2011,Tabik-Optimal_tilt_and_orientation_maps_a_multi_algorithm_approach_for_heterogeneous_multicore_GPU_systems:2013},
i.e., no network communication is required, or only a few processes
deployed over a handful of nodes \cite{Huang-Explorations_of_the_implementation_of_a_parallel_IDW_algorithm_in_a_Linux_cluster:2011}.
Specifically, we are referring to network saturation and idle processes
within the master-worker model. Generally speaking, a single communicating
process, e.g., the master, is usually not able to saturate the network
connection of a node. Using more than one MPI process per node might
solve this problem, but possible rank-ordering problems may appear,
thus restricting the full utilization of the network \cite{Rabenseifner-Hybrid_MPI_OpenMP_parallel_programming_on_clusters_of_multicore_nodes:2009}.
Another issue appears when the master process executes the MPI code,
in which case other processes sleep, making a serial use of the communication
component of the system. Consequently, the master process becomes
the bottleneck of the parallel implementation as the number of worker
processes it has to serve grows. This situation is also common when
dealing with the metadata of a spatial region, which may relate to
several elements of a RSG, making it a frequent cause of load imbalance
\cite{Gong_Parallel_agent_based_simulation_of_individual_level_spatial_interactions_within_a_multicore_computing_environment:2012,Hawick_Distributed_frameworks_and_parallel_algorithms_for_processing_large_scala_geographic_data:2003,Widener_Developing_a_parallel_computational_implementation_of_AMOEBA:2012}.
In our case, the transmitter configuration and its antenna diagram
represent metadata that are complementary to the sub-region that a
transmitter covers.

Hybrid MPI-OpenMP implementations \cite{Tabik-High_performance_three_horizon_composition_algorithm_for_large_scale_terrains:2011,Tabik-Optimal_tilt_and_orientation_maps_a_multi_algorithm_approach_for_heterogeneous_multicore_GPU_systems:2013},
in which no MPI calls are issued inside the OpenMP-parallel regions,
also fail to saturate the network \cite{Rabenseifner-Hybrid_MPI_OpenMP_parallel_programming_on_clusters_of_multicore_nodes:2009}.
A possible solution to this problem is to improve the communication
overlap among the processes. To this end, we have implemented non-blocking
point-to-point MPI operations, and an independent thread in the worker
process to save the intermediate results to a DB. We use one such
database system per computer cluster, which also serves the input
data to the GRASS, in order to aggregate the partial results of the
path-loss predictions or to visualize them. It is important to note
that any kind of database system may be used. By this we mean relational,
distributed~\cite{Ozsu_Principles_of_distributed_database_systems:2011}
or even those of the NoSQL type~\cite{Stonebraker_SQL_databases_vs_NoSQL_databases:2010}.
Nevertheless, in this study we use a central relational database system,
since they are the most popular and widely available ones. Additionally,
the non-blocking message-passing technique used to distribute the
work-load among the nodes provides support for heterogeneous environments.
As a result, computing nodes featuring more capable hardware receive
more work than those with weaker configurations, thus ensuring a better
utilization of the available computing resources despite hardware
diversity and improved load balancing.

\subsubsection{Master process\label{sub:Master-process}}

\begin{figure*}[tbh]
\begin{minipage}[t]{0.45\textwidth}%
\centering

\includegraphics[width=0.48\columnwidth]{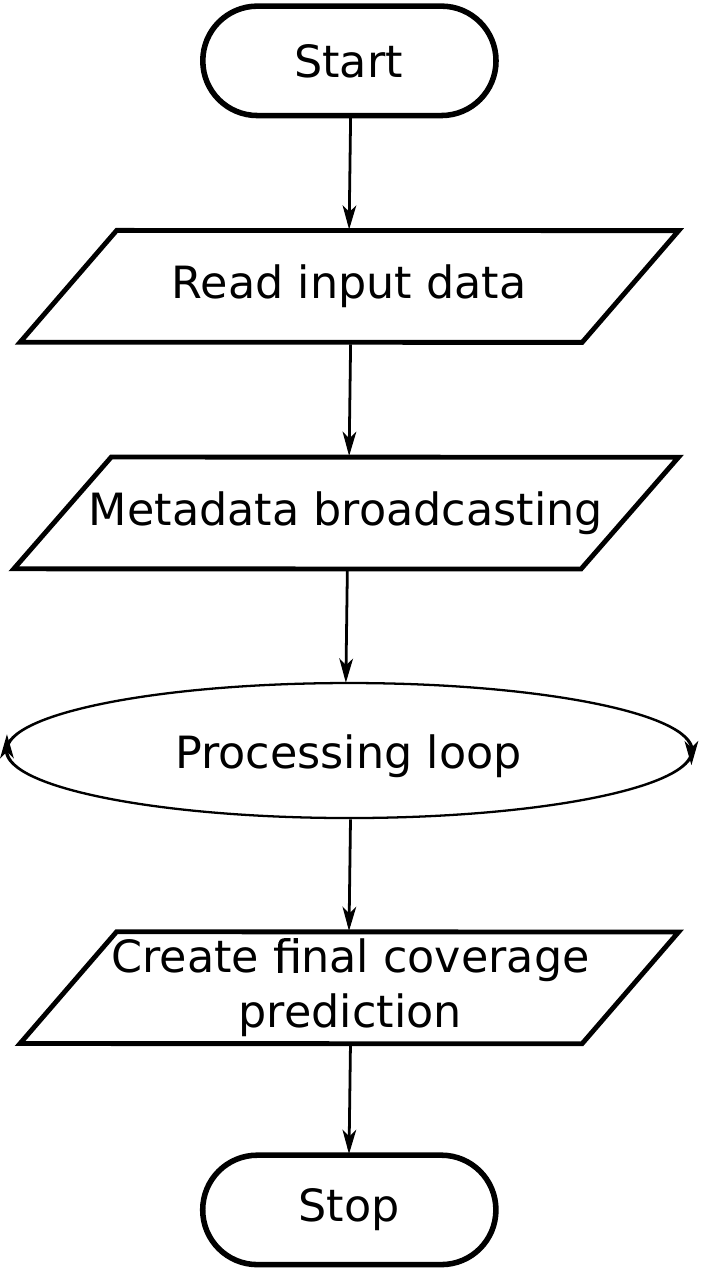}

\caption{Flow diagram of the master process.\label{fig:master_process}}
\end{minipage}\hfill{}%
\begin{minipage}[t]{0.45\textwidth}%
\centering

\includegraphics[width=1\columnwidth]{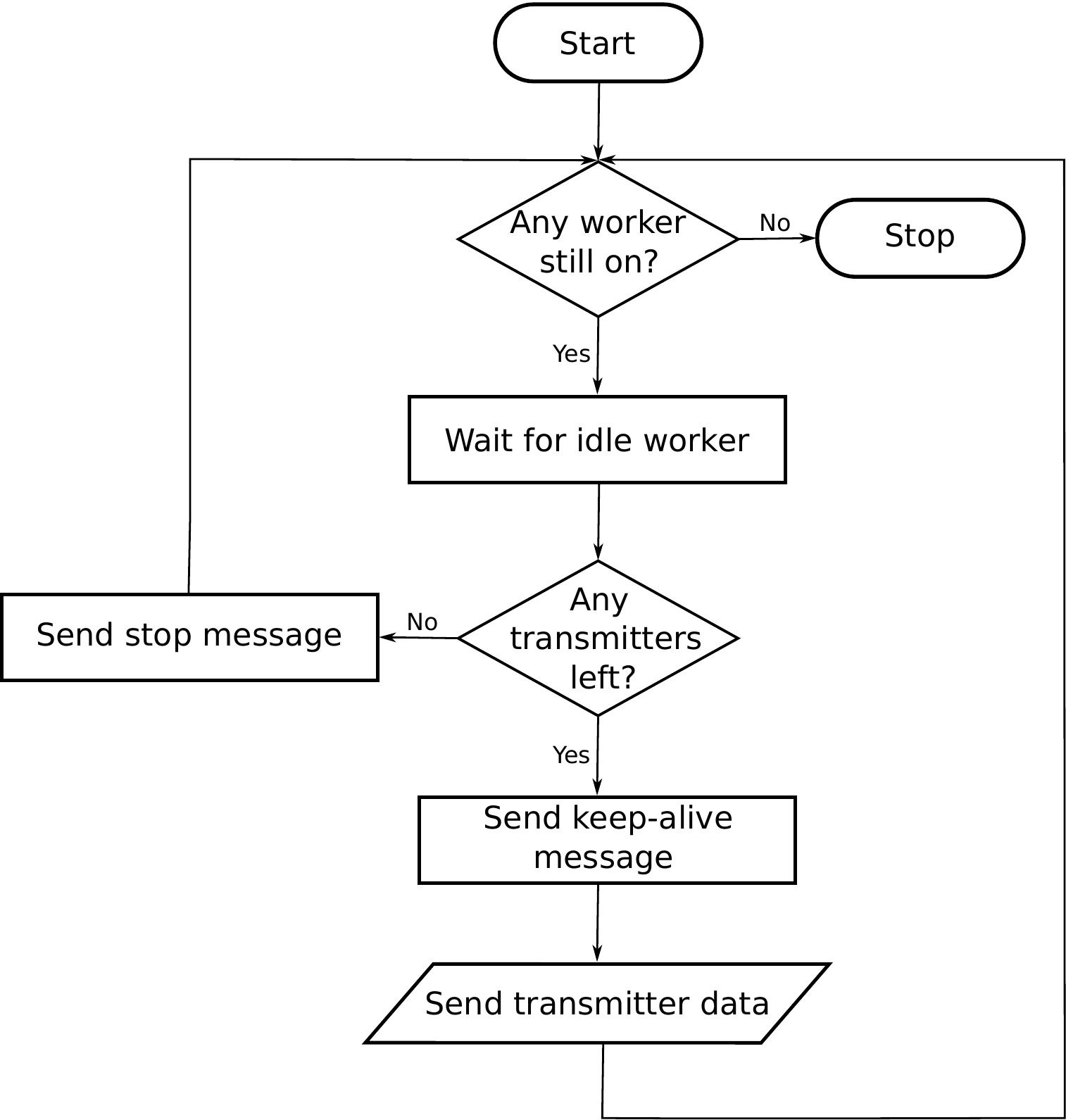}

\caption{Flow diagram of the ``Processing loop'' step of the
master process.\label{fig:processing_loop_in_master_process}}
\end{minipage}
\end{figure*}

The master process, for which the flow diagram is given in \prettyref{fig:master_process},
is the only component that runs within the GRASS environment. As soon
as the master process starts, the input parameters are read. This
step corresponds to ``Read input data'' in \prettyref{fig:master_process},
and it is carried out in a similar way as in the serial version. The
next step delivers the metadata that is common to all the transmitters
and the whole region to all the processes (see ``Metadata broadcasting''
in \prettyref{fig:master_process}). Before distributing the work
among the worker processes, the master process proceeds to decompose
the loaded raster data into 2D matrices of basic-data-type elements,
e.g., \emph{float} or \emph{double}, before dispatching them to the
multiple worker processes. In this case, the decomposition applies
to the DEM and the clutter data only, but it could be applied to any
point-based data set. In the next step, the master process starts
an asynchronous message-driven processing loop (see ``Processing
loop'' in \prettyref{fig:master_process}), the main task of which
is to assign and distribute the sub-region and configuration data
of different transmitters among the idle worker processes.

The flow diagram shown in \prettyref{fig:processing_loop_in_master_process}
illustrates the ``Processing loop'' step of the master process.
In the processing loop, the master process starts by checking the
available worker processes, which will calculate the radio-coverage
prediction for the next transmitter. It is worth pointing out that
this step also serves as a stopping condition for the processing loop
itself (see ``Any worker still on?'' in \prettyref{fig:processing_loop_in_master_process}).
The active worker processes inform the master process that they are
ready to compute by sending an idle message (see ``Wait for idle
worker'' in \prettyref{fig:processing_loop_in_master_process}).
The master process then announces to the idle worker process that
it is about to receive new data for the next calculation, and it dispatches
the complete configuration of the transmitter to be processed (see
``Send keep-alive message'' and ``Send transmitter data'' steps,
respectively, in \prettyref{fig:processing_loop_in_master_process}).
This is only done in the case that there are transmitters for which
the coverage prediction has yet to be calculated (see ``Any transmitters
left?'' in \prettyref{fig:processing_loop_in_master_process}). The
processing loop of the master process continues to distribute the
transmitter data among the worker processes, which asynchronously
become idle as they finish the radio-prediction calculations they
have been assigned by the master process. When there are no more transmitters
left, all the worker processes announcing they are idle will receive
a shutdown message from the master process, indicating to them that
they should stop running (see ``Send stop message'' in \prettyref{fig:processing_loop_in_master_process}).
The master process will keep doing this until all the worker processes
have finished (see ``Any worker still on?'' in \prettyref{fig:processing_loop_in_master_process}),
thus fulfilling the stopping condition for the processing loop.

Finally, the last step of the master process is devoted to creating
the final output of the calculation, e.g., a raster map (see ``Create
final coverage prediction'' in \prettyref{fig:master_process}).
The final coverage prediction of all the transmitters is an aggregation
from the individual path-loss results created by each of the worker
processes during the ``Processing loop'' phase in \prettyref{fig:master_process},
which provides the source data for the final raster map. The aggregation
of the individual transmitter path-loss results is accomplished by
issuing an SQL query over the database tables containing the partial
results, in a similar way as in the serial version.

\subsubsection{Worker processes}

An essential characteristic of the worker processes is that they are
completely independent of the GRASS, i.e., they do not have to run
within the GRASS environment nor use any of the GRASS libraries to
work. This aspect significantly simplifies the deployment phase to
run PRATO on a computer cluster, since no GRASS installation is needed
on the computing nodes hosting the worker processes.

One possibility to overcome the thread-safety limitation of the GRASS
is to save the transmitter path-loss predictions through the master
process, thus avoiding concurrent access. However, for the workers
to send intermediate results back to the master process, e.g., as
in \cite{Akhter-GRASS_GIS_on_high_performance_computing_with_MPI_OpenMP_and_Ninf-G:2010,Huang-Explorations_of_the_implementation_of_a_parallel_IDW_algorithm_in_a_Linux_cluster:2011},
is a major bottleneck for the scalability of a parallel implementation.
The scalability is limited by the master process, because it must
serially process the received results in order to avoid inconsistencies
due to concurrent access. Instead, our approach allows each of the
worker processes to output its intermediate results into a DB, i.e.,
each path-loss prediction in its own table. Additionally, worker processes
do this from an independent thread, which runs concurrently with the
calculation of the next transmitter received from the master process.
In this way, the overlap between the calculation and communication
significantly hides the latency created by the result-dumping task,
thus making better use of the available system resources.

\begin{figure*}[tbh]
\begin{minipage}[t]{0.52\textwidth}%
\centering

\includegraphics[width=0.85\columnwidth]{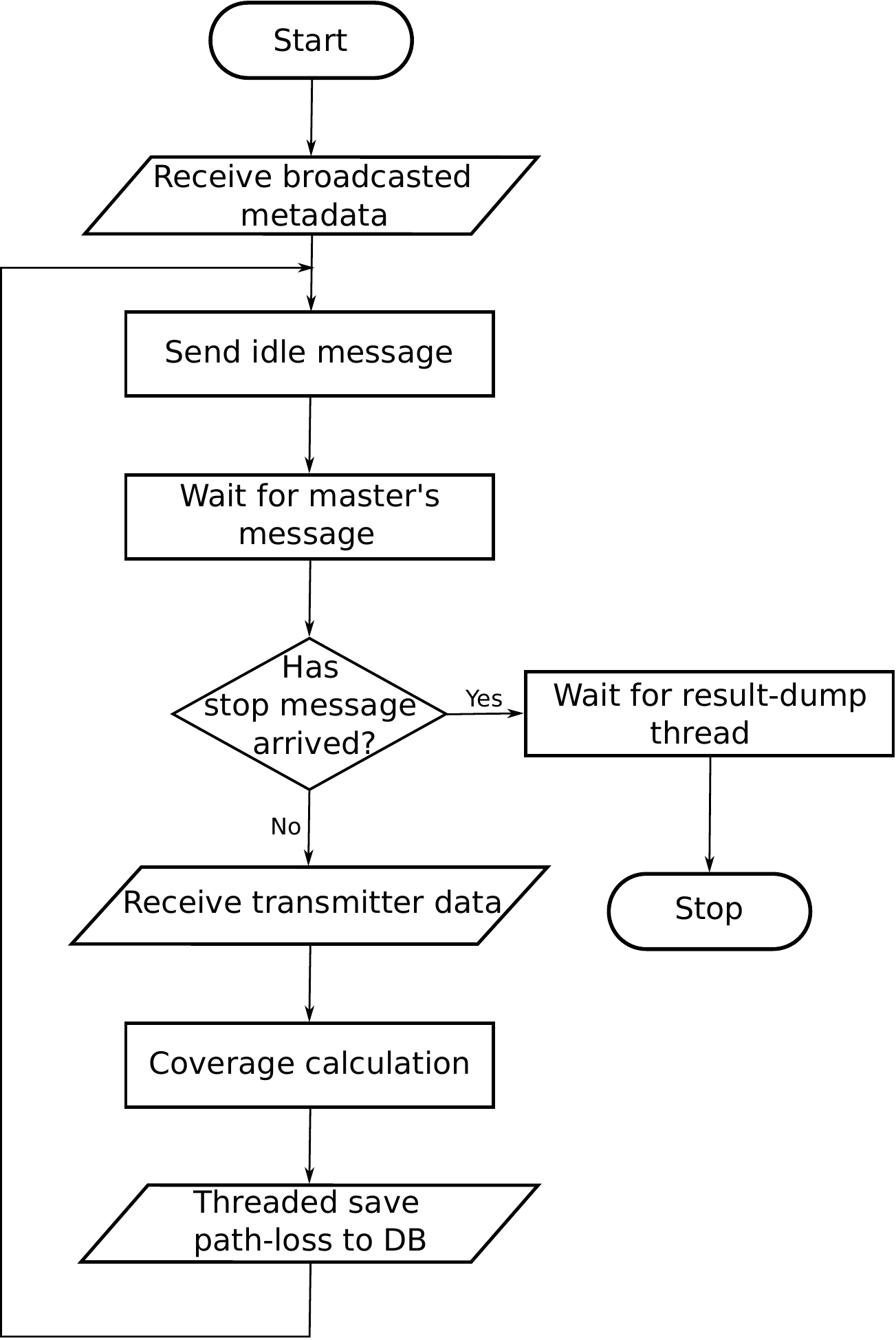}

\caption{Flow diagram of a worker process.\label{fig:worker_process_flow_diagram}}
\end{minipage}\hfill{}%
\begin{minipage}[t]{0.47\textwidth}%
\centering

\includegraphics[width=1\columnwidth]{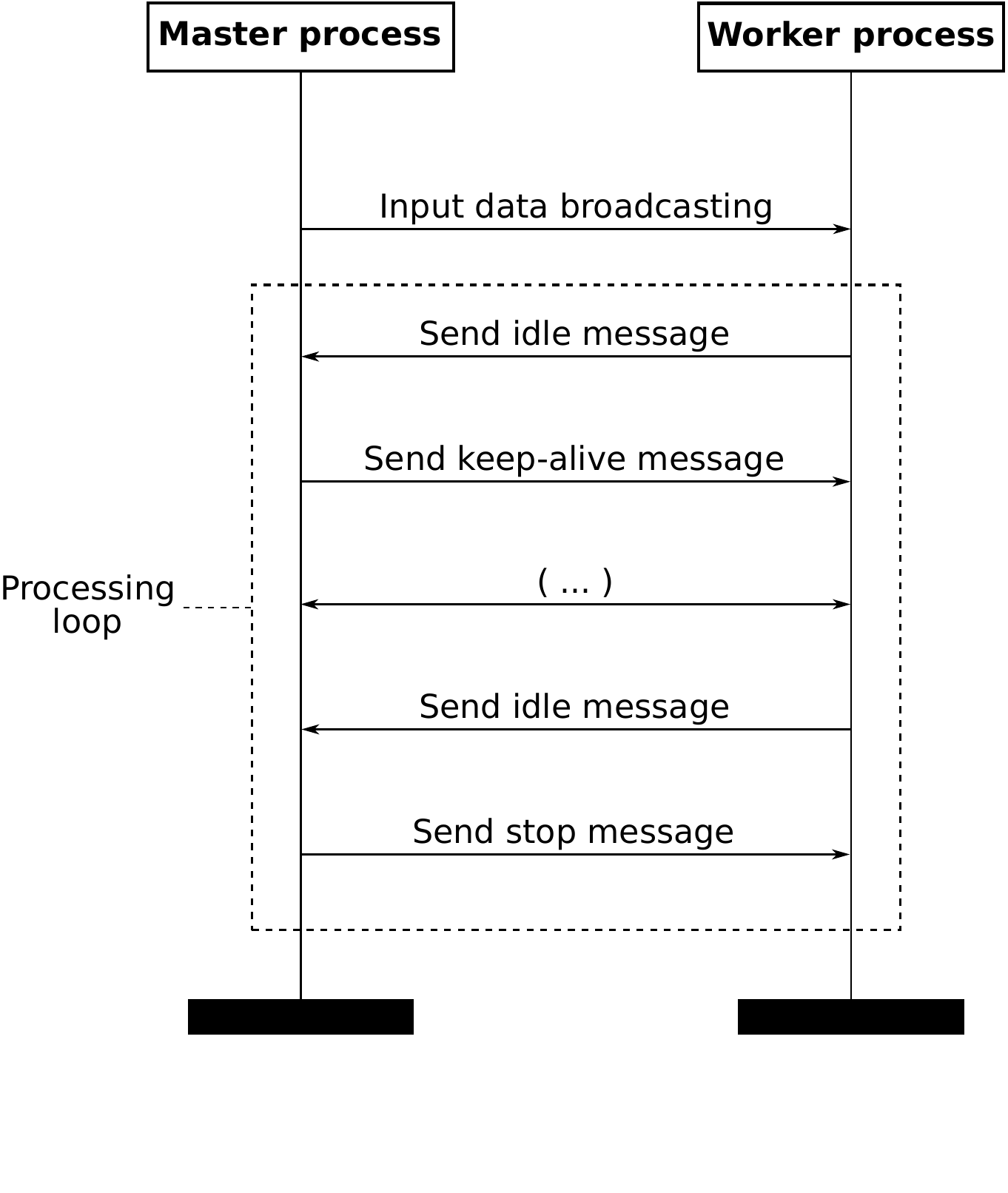}

\caption{Communication diagram, showing message passing between
master and one worker process.\label{fig:master_worker_communication}}
\end{minipage}
\end{figure*}

The computations of the worker processes, for which the flow diagram
is given in \prettyref{fig:worker_process_flow_diagram}, begin by
receiving metadata about the transmitters and the geographical area
from the master process during the initialization time (see ``Receive
broadcasted metadata'' in \prettyref{fig:worker_process_flow_diagram}).

After the broadcasted metadata are received by all the worker processes,
each one proceeds to inform the master process that it is ready (i.e.,
in an idle state) to receive the transmitter-configuration data that
defines which transmitter path-loss prediction to perform (see ``Send
idle message'' in \prettyref{fig:worker_process_flow_diagram}).
If the master process does not give the instruction to stop processing
(see ``Has stop message arrived?'' in \prettyref{fig:worker_process_flow_diagram}),
the worker process collects the sub-region spatial data and the transmitter
configuration (see ``Receive transmitter data'' in \prettyref{fig:worker_process_flow_diagram}).
In the event that a stop message is received, the worker process will
wait for any result-dumping thread to finish (see ``Wait for result-dump
thread'' in \prettyref{fig:worker_process_flow_diagram}) before
shutting down. The coverage calculation itself follows a similar design
as the serial version (see ``Coverage calculation'' in \prettyref{fig:worker_process_flow_diagram}).

As mentioned before, the worker process launches an independent thread
to save the path-loss prediction of the target transmitter to a database
table (see ``Threaded save path-loss to DB'' in \prettyref{fig:worker_process_flow_diagram}).
It is important to note that there is no possibility of data inconsistency
due to the saving task being executed inside a thread, since path-loss
data from different workers belong to different transmitters and are,
at this point of the process, mutually exclusive.

\subsubsection{Master-worker communication\label{sub:Master-worker-communication}}

Similar to \cite{Tabik-High_performance_three_horizon_composition_algorithm_for_large_scale_terrains:2011,Tabik-Optimal_tilt_and_orientation_maps_a_multi_algorithm_approach_for_heterogeneous_multicore_GPU_systems:2013},
the message-passing technique used in this work enables a better use
of the available computing resources, both in terms of scalability
and load balancing, while introducing a negligible overhead. This
last point is supported by the experimental results, introduced in
Section~\ref{sub:Strong-scalability}.

The first reason to implement the message-passing technique is to
support heterogeneous computing environments. In particular, our approach
focuses on taking full advantage of the hardware of each computing
node, thus explicitly avoiding the bottlenecks introduced by the slowest
computing node in the cluster. This problem appears when evenly distributing
the data among the worker processes on disparate hardware, as in~\cite{Akhter_Porting_GRASS_raster_module_to_distributed_computing:2007,Huang-Explorations_of_the_implementation_of_a_parallel_IDW_algorithm_in_a_Linux_cluster:2011},
being more noticeable with a larger number of computing nodes and
processes. In other words, computing nodes that deliver better performance
have more calculations assigned to them. Moreover, in real-world scenarios,
it is often the case that a large number of dedicated computing nodes
featuring exactly the same configuration is difficult to find, i.e.,
not every organization owns a computer cluster.

A second reason for selecting a message-passing technique is related
to the flexibility it provides for load balancing, which is of greater
importance when dealing with extra data or information besides just
spatial data \cite{Hawick_Distributed_frameworks_and_parallel_algorithms_for_processing_large_scala_geographic_data:2003}.
This can be seen in \prettyref{fig:processing_loop_in_master_process},
where the master process, before delivering the spatial subset and
transmitter-configuration data, sends a message to the worker process,
indicating that it is about to receive more work. This a priori meaningless
message plays a key role in correctly supporting the asynchronous
process communication. Notice that the subset of spatial data that
a worker process receives is directly related to the transmitter for
which the prediction will be calculated. Similar to~\cite{Tabik-High_performance_three_horizon_composition_algorithm_for_large_scale_terrains:2011,Tabik-Optimal_tilt_and_orientation_maps_a_multi_algorithm_approach_for_heterogeneous_multicore_GPU_systems:2013},
this problem-specific property enables the use of a data-decomposition
technique based on a block partition of spatial data, e.g., the DEM
and clutter data.

In general, there are many different ways a parallel program can be
executed, because the steps from the different processes can be interleaved
in various ways and a process can make non-deterministic choices \cite{Siegel_Verification_of_halting_properties_for_MPI_programs:2007},
which may lead to situations such as race conditions \cite{Clemencon_MPI_Race_detection:1995}
and deadlocks. A deadlock occurs whenever two or more running processes
are waiting for each other to finish, and thus neither ever does.
To prevent PRATO from deadlocking, message sending and receiving should
be paired, i.e., an equal number of send and receive messages on the
master and worker sides \cite{Siegel_Verification_of_halting_properties_for_MPI_programs:2007}.

\prettyref{fig:master_worker_communication} depicts the master-worker
message passing, from which the transmitter-data transmission has
been excluded for clarity. Notice how each idle message sent from
the worker process is paired with an answer from the master process,
whether it is a keep-alive or a stop message.

\section{Simulations \label{sec:Simulations}}

Considering the large computational power needed for predicting the
radio-coverage of a real mobile network, the use of a computer cluster
is recommended. A computer cluster is a group of interconnected computers
that work together as a single system. Computer clusters typically
consist of several commodity PCs connected through a high-speed local-area
network (LAN) with a distributed file system, like NFS~\cite{Shepler_Network_file_system:2003}.
One such system is the DEGIMA cluster~\cite{Hamada_Cluster_of_GPUs:2010}
at the Nagasaki Advanced Computing Center (NACC) of the Nagasaki University
in Japan. This system ranked in the TOP 500 list of supercomputers
until June 2012%
\footnote{http://www.top500.org%
}, and in June 2011 it held third place in the Green 500 list%
\footnote{http://www.green500.org%
} as one of the most energy-efficient supercomputers in the world.

This section presents the simulations and analyses of the parallel
version of PRATO. Our aim is to provide an exhaustive analysis of
the performance and scalability of the parallel implementation in
order to achieve the objectives of this work. The most common usage
case for PRATO is to perform a radio-coverage prediction for multiple
transmitters. Therefore, a straight-forward parallel decomposition
is to divide a given problem instance by transmitter, for which each
coverage prediction is calculated by a separate worker process.

The following simulations were carried out on 34 computing nodes of
the DEGIMA cluster. The computing nodes are connected by a LAN, over
a Gigabit Ethernet interconnect. As mentioned before, the reason for
using a high-end computer cluster such as DEGIMA is to explore by
experimentation the advantages and drawbacks of the introduced methods.
However, this does not imply any loss of generality when applying
these principles over a different group of networked computers, i.e.,
not acting as a computer cluster.

Each computing node of DEGIMA features one of two possible configurations,
namely:
\begin{itemize}
\item Intel Core i5-2500T quad-core processor CPU, clocked at 2.30 GHz,
with 16 GB of RAM; and
\item Intel Core i7-2600K quad-core processor CPU, clocked at 3.40 GHz,
also with 16 GB of RAM.
\end{itemize}
During the simulation runs, the nodes equipped with the Intel i5 CPU
host the worker processes, whereas the master process and the PostgreSQL
database server (version 9.1.4) each run on a different computing
node, featuring an Intel i7 CPU. The database server performs all
its I/O operations on the local file system, which is mounted on an
8~GB RAM disk. During the simulations, the path-loss predictions
of 5,120 transmitters occupied less than 4~GB of this partition.

All the nodes are equipped with a Linux 64-bit operating system (Fedora
distribution). As the message passing implementation we use OpenMPI,
version 1.6.1, which has been manually compiled with the distribution-supplied
gcc compiler, version 4.4.4.

\subsection{Test networks}

To test the parallel performance of PRATO, we prepared different problem
instances that emulate real radio networks of different sizes. In
order to create the synthetic test data-sets with an arbitrary number
of transmitters, we used the real data of a group of 2,000 transmitters,
which we randomly replicate and distribute over the whole target area.
The configuration parameters of these 2,000 transmitters were taken
from the LTE network deployed in Slovenia by Telekom Slovenije, d.d.
The path-loss predictions were calculated using the Walfisch-Ikegami
model. The digital elevation model has an area of 20,270~km$^{2}$,
with a resolution of 25~m$^{2}$. The clutter data extends over the
same area and resolution, containing different levels of signal loss
due to land usage. For all the points within a radius of 20~km around
each transmitter, we assume that the receiver is positioned 1.5~m
above the ground, and the frequency is set to 1,843~MHz.

\subsection{Weak scalability}

This set of simulations is meant to analyze the scalability of the
parallel implementation in cases where the workload assigned to each
process (one MPI process per processor core) remains constant as we
increase the number of processor cores and the total size of the problem,
i.e., the number of transmitters deployed over the target area is
directly proportional to the number of processor cores and worker
processes. We do this by assigning a constant number of transmitters
per core, while increasing the number of cores hosting the worker
processes. Here we test for the following numbers of transmitters
per worker/core: $\{5,10,20,40,80\}$, by progressively doubling the
number of worker processes from 1 to 64.

Problems that are particularly well-suited for parallel computing
exhibit computational costs that are linearly dependent on the size
of the problem. This property, also referred to as algorithmic scalability,
means that proportionally increasing both the problem size and the
number of cores results in a roughly constant time to solution.

The master-worker (MW) configuration performs result aggregation continuously,
i.e., while receiving the intermediate results from the worker processes.
In contrast, the master-worker-DB (MWD) setup performs the result
aggregation as the final step. With this set of experiments, we would
like to investigate how the proposed MWD technique compares with the
classic MW approach in terms of scalability when dealing with different
problem instances and numbers of cores.

An important fact about the presented simulations when using multi-threaded
implementations is to avoid oversubscribing a computing node. For
example, if deploying four worker processes over a quad-core CPU,
the extra threads will have a counter effect on the parallel efficiency,
since the CPU resources would be exhausted, which slows the whole
process down. For this reason, we have deployed three worker processes
per computing node, leaving one core free for executing the extra
threads.

\subsubsection{Results and discussion}

\begin{figure}[tbh]
\centering

\includegraphics[width=1\columnwidth]{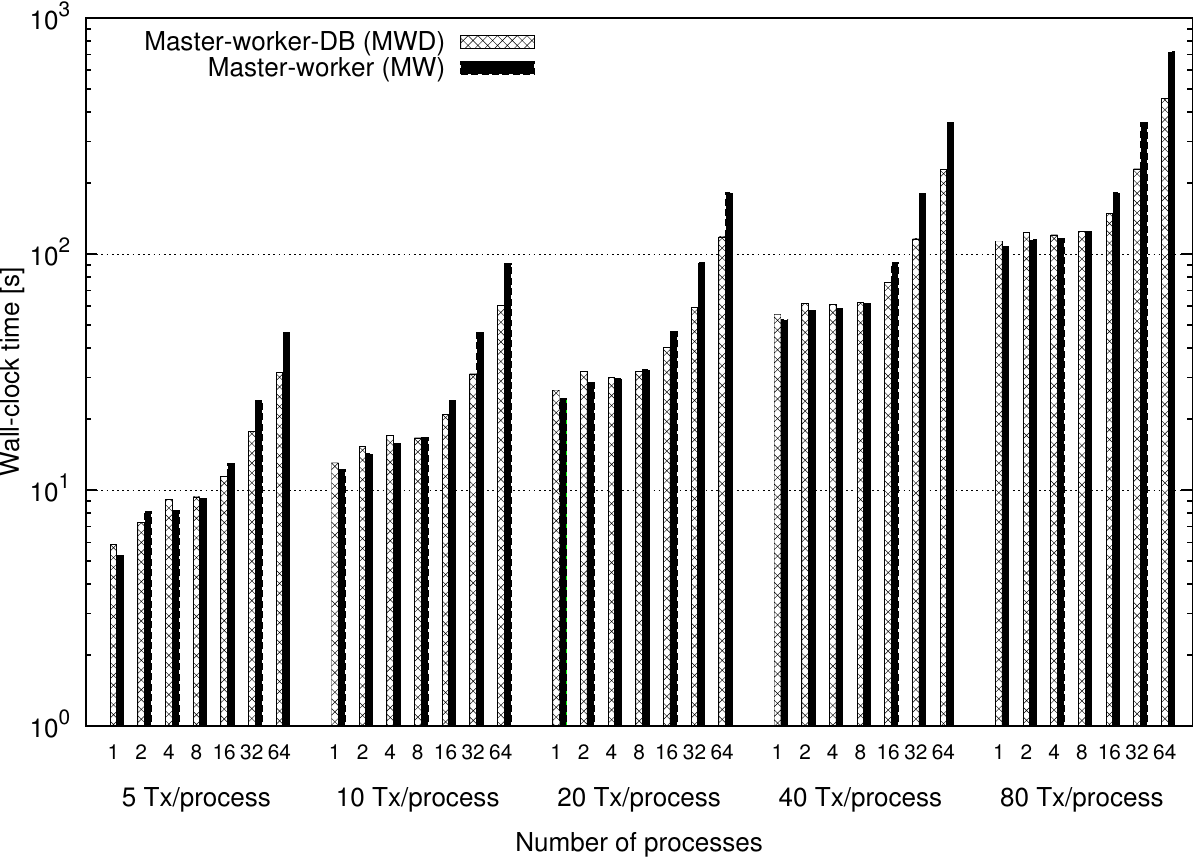}

\caption{Measured wall-clock time for weak-scalability experiments,
featuring MW and MWD setups. Experiments allocate one MPI worker process per core.
The wall-clock time axis is expressed in a base-10 logarithmic scale, whereas the axis representing
the number of cores is expressed in a base-2 logarithmic scale.\label{fig:weak_scalability_time}}
\end{figure}

The results represent the best running time out of a set of 20 independent
simulation runs, in which we randomly selected both the transmitters
and the rank ordering of the worker processes. The collected running
times for the weak-scalability experiments are shown in \prettyref{fig:weak_scalability_time}.
All the measurements express wall-clock times in seconds for each
setup and problem instance, defined as the number of transmitters
per process (TX/process). The wall-clock time represents the real
time that elapses from the start of the master process to its end,
including the time that passes while waiting for the resources to
become available. The running-time improvements of the master-worker-DB
against the master-worker setup are shown in Table \ref{tab:weak_scaling-time_gain}.

\begin{table}[tbh]
\centering

\caption{Running-time gain (in percent) of the simulations for the weak-scalability
of the MWD setup relative to the classic MW approach.\label{tab:weak_scaling-time_gain}}

{\footnotesize{}}%
\begin{tabular}{cccccccc}
\cmidrule{2-8} 
 & \multicolumn{7}{c}{{\footnotesize{Number of cores}}}\tabularnewline\addlinespace
\midrule 
{\footnotesize{TX/core}} & {\footnotesize{1}} & {\footnotesize{2}} & {\footnotesize{4}} & {\footnotesize{8}} & {\footnotesize{16}} & {\footnotesize{32}} & {\footnotesize{64}}\tabularnewline
\midrule
{\footnotesize{5}} & {\footnotesize{-11.39}} & {\footnotesize{-10.42}} & {\footnotesize{-11.14}} & {\footnotesize{-0.95}} & {\footnotesize{11.75}} & {\footnotesize{26.15}} & {\footnotesize{32.53}}\tabularnewline
{\footnotesize{10}} & {\footnotesize{-5.84}} & {\footnotesize{-7.78}} & {\footnotesize{-7.67}} & {\footnotesize{0.91}} & {\footnotesize{12.81}} & {\footnotesize{33.28}} & {\footnotesize{33.55}}\tabularnewline
{\footnotesize{20}} & {\footnotesize{-8.59}} & {\footnotesize{-10.88}} & {\footnotesize{-1.04}} & {\footnotesize{1.95}} & {\footnotesize{14.29}} & {\footnotesize{35.23}} & {\footnotesize{35.27}}\tabularnewline
{\footnotesize{40}} & {\footnotesize{-5.26}} & {\footnotesize{-6.90}} & {\footnotesize{-3.68}} & {\footnotesize{-0.67}} & {\footnotesize{17.27}} & {\footnotesize{36.23}} & {\footnotesize{36.65}}\tabularnewline
{\footnotesize{80}} & {\footnotesize{-5.29}} & {\footnotesize{-7.11}} & {\footnotesize{-3.20}} & {\footnotesize{-0.31}} & {\footnotesize{17.94}} & {\footnotesize{36.32}} & {\footnotesize{36.57}}\tabularnewline
\bottomrule
\end{tabular}
\end{table}

The time measurements observed from the weak-scalability results show
that the classic MW approach performs well for up to four worker processes.
When using eight worker processes, the MW setup is practically equivalent
to the MWD approach, indicating that the master process is being fully
exploited. When increasing the problem size and the number of worker
processes to 16, the running-time gain is already clear, favoring
the MWD configuration. This gain keeps growing, although slower, as
we increase the number of worker processes to 32 and 64, confirming
our hypothesis that in a classic MW approach, the parallel efficiency
is bounded by the capacity of the master process to serve an increasing
number of worker processes. Interestingly, the gain when using 32
and 64 worker processes is almost the same. After further investigation,
we found the reason for this behavior was due to the LAN being completely
saturated by the worker processes. Consequently, they have to wait
for the network resources to become available before sending or receiving
data, which is not the case when running the MW setup. Therefore,
using the MWD approach we hit a hardware constraint, meaning that
the bottleneck is no longer at the implementation level. Moreover,
since the master process is far from overloaded when serving 64 worker
processes, we can expect the MWD approach will keep scaling if we
use a faster network infrastructure, e.g., 10-gigabit Ethernet or
InfiniBand.

Certainly, the parallel version of PRATO, when using the MWD approach,
scales better when challenged with a large number of transmitters
(5,120 for the biggest instance) over 64 cores. This fact shows PRATO
would be able to calculate the radio-coverage prediction for real
networks in a feasible amount of time, since many operational radio
networks have already deployed a comparable number of transmitters,
e.g., the 3G network within the Greater London Authority area, in
the UK \cite{Number_of_base_stations_in_England}.

Not being able to achieve perfect weak scalability using the MWD setup
is due to a number of factors. Specifically, the overhead time of
the serial sections of the parallel process grow proportionally with
the number of cores, e.g., aggregation of the intermediate results,
although the total contribution of this overhead remains low for large
problem sizes. Moreover, the communication overhead grows linearly
with the number of cores used. Consequently, we can confirm the findings
of \cite{Huang-Explorations_of_the_implementation_of_a_parallel_IDW_algorithm_in_a_Linux_cluster:2011},
who concluded that the data-set size should be large enough for the
communication overhead to be hidden by the calculation time, for parallelization
to be profitable in terms of a running-time reduction.

\subsection{Strong scalability\label{sub:Strong-scalability}}

This set of simulations is meant to analyze the impact of increasing
the number of computing cores for a given problem size, i.e., the
number of transmitters deployed over the target area does not change,
while only the number of worker processes used is increased. Here
we test for the following number of transmitters: \{1,280, 2,560,
5,120\}, by gradually doubling the number of workers from 1 to 64
for each problem size.

\subsubsection{Results and discussion}

\begin{figure}[tbh]
\centering

\includegraphics[width=1\columnwidth]{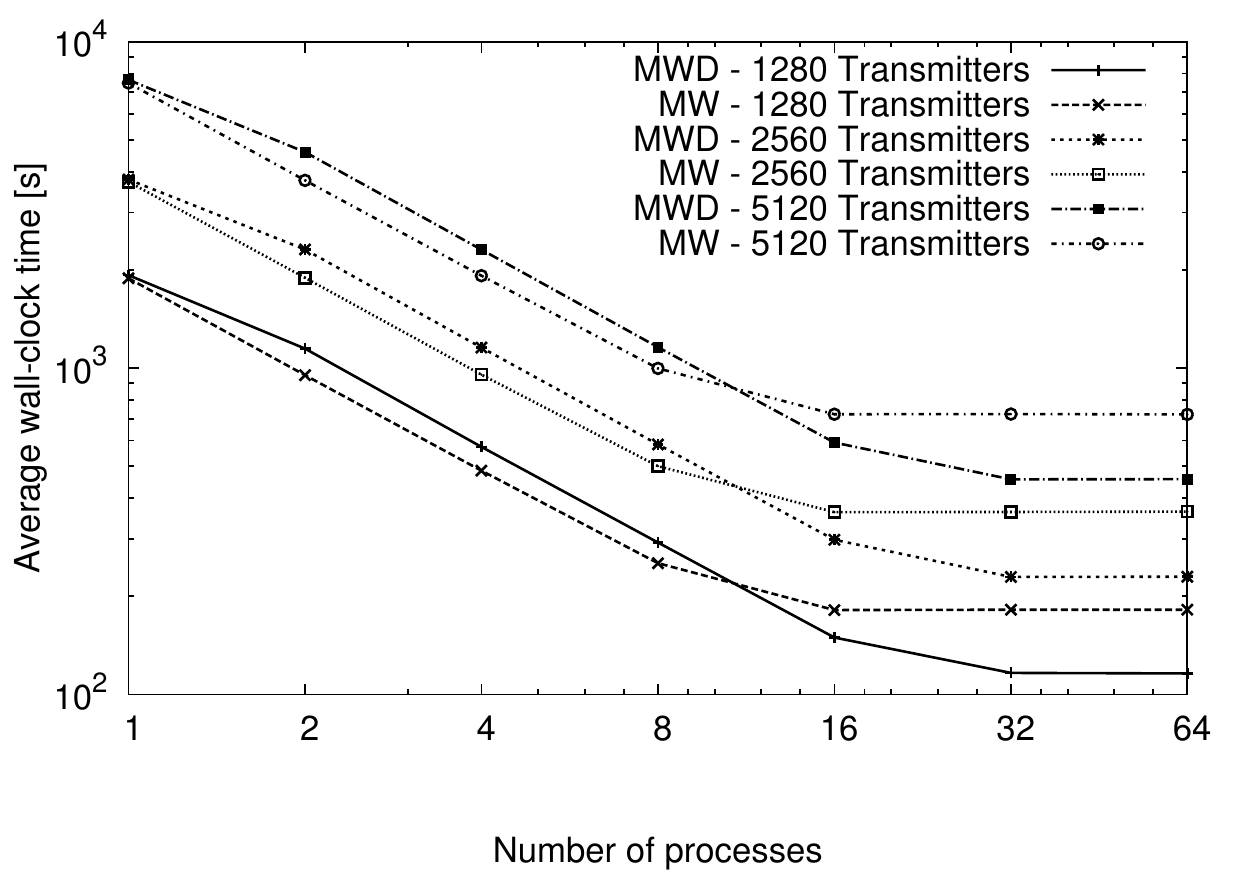}

\caption{Measured wall-clock time for strong-scalability experiments,
featuring MW and MWD setups. Experiments assigned one MPI worker process per core. 
The wall-clock time axis is expressed in a base-10 logarithmic scale, whereas the axis representing
the number of cores is expressed in a base-2 logarithmic scale.\label{fig:strong_scalability_time}}
\end{figure}

Similar to the weak-scalability experiments, these time measurements
show that when applying a classic MW approach the running-time reduction 
starts flattening when more than eight worker processes
are used. Moreover, the running times for 16, 32 and 64 worker processes
are the same, i.e., it does not improve due to the master process
being saturated. In contrast, when using our MWD technique,
the running-time reduction improves for up to 32 worker processes,
after which there is no further improvement since the network is being
fully exploited. These results clearly show that when applying parallelization
using a larger number of worker processes, the master process becomes
the bottleneck of the MW approach. When using the MWD configuration,
a steady running-time reduction is observed, until a hardware constraint
is hit, e.g., the network infrastructure.

We have also measured the overhead of sending/receiving asynchronous
messages in order to support heterogeneous systems, which is lower
than 0.02\% of the total running time for the MW experiments, and
0.01\% for the MWD experimental set.

In order to further analyze how well the application scales using
the MW and MWD approaches, we measured the performance of the parallel
implementation in terms of its speedup, which is defined as

\begin{equation}
S(NP)=\frac{execution\, time\, for\, base\, case}{execution\, time\, for\, NP\, cores},\label{eq:speedup}
\end{equation}

\noindent where $NP$ is the number of cores executing the worker
processes. As the base case for comparisons we chose the parallel
implementation running on only one core and decided against using
the serial implementation. We consider that the serial implementation
is not a good base comparison for the parallel results as it does
not reuse the resources between each transmitter-coverage calculation
and it does not overlap the I/O operations with the transmitter computations.
In practice, this means that several concatenated runs of the serial
version would be considerably slower than the parallel but single
worker implementation.

\begin{figure}[tbh]
\begin{minipage}[t]{0.48\textwidth}%
\centering

\includegraphics[width=1\columnwidth]{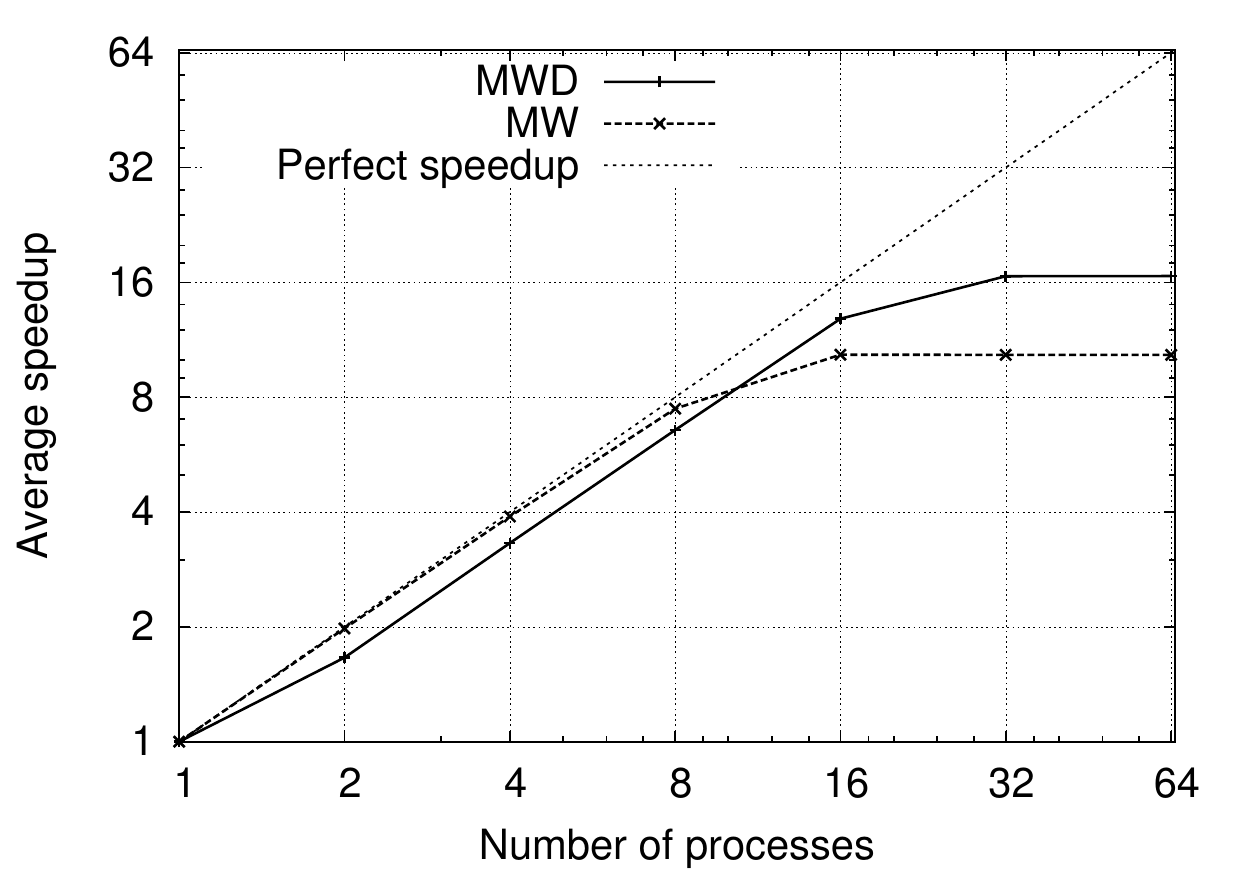}

\caption{Average speedup for strong-scalability experiments. 
The speedup axis is expressed in a base-2 logarithmic
scale, and the axis representing the number of cores is expressed
in a base-2 logarithmic scale.\label{fig:strong_scalability_speedup}}
\end{minipage}\hfill{}%
\begin{minipage}[t]{0.48\textwidth}%
\centering

\includegraphics[width=1\columnwidth]{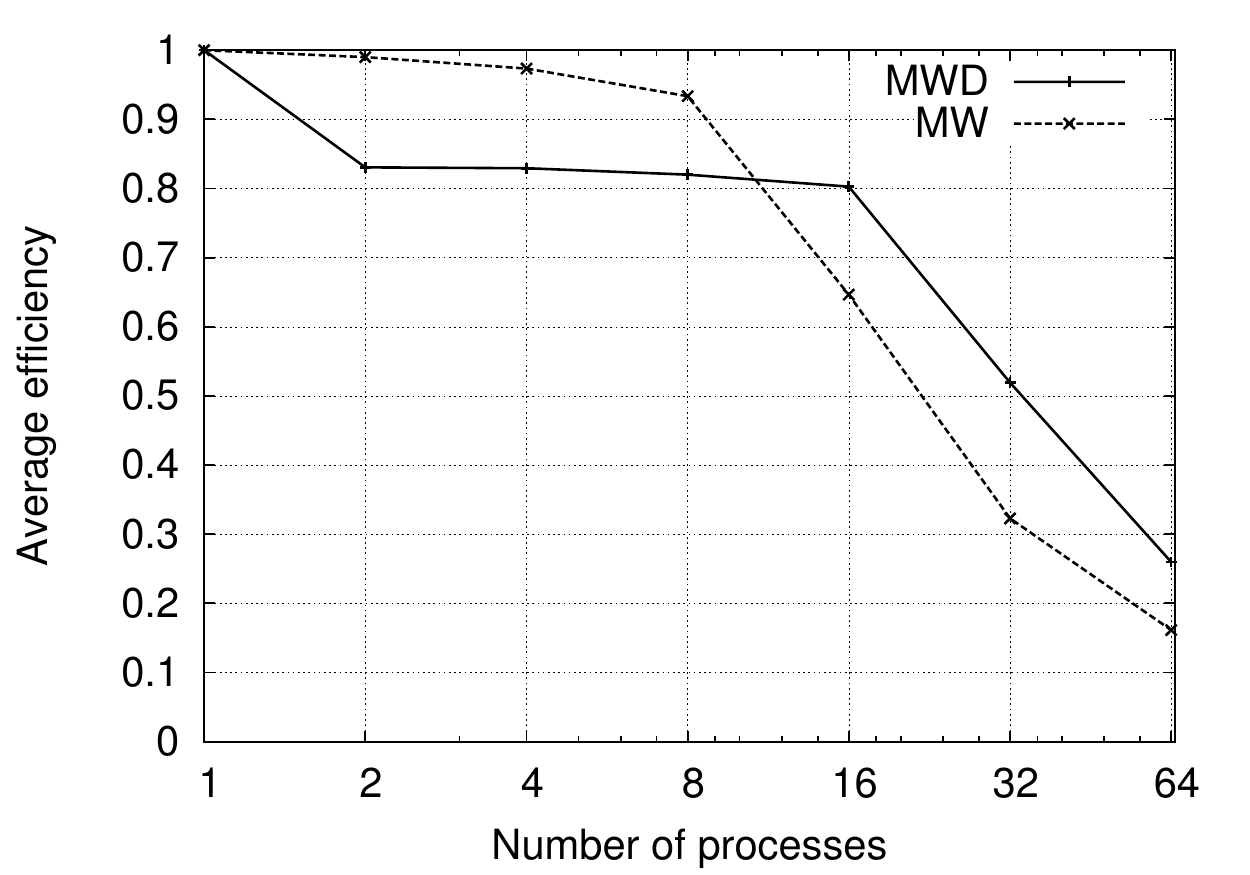}

\caption{Average parallel efficiency for strong-scalability experiments.
The parallel-efficiency axis is expressed in a linear
scale, whereas the axis representing the number of cores is expressed
in a base-2 logarithmic scale.\label{fig:strong_scalability_efficiency}}
\end{minipage}
\end{figure}

Using the speedup metric, linear scaling is achieved when the obtained
speedup is equal to the total number of processors used. However,
it should be noted that a perfect speedup is almost never achieved,
due to the existence of serial stages within an algorithm and the
communication overhead of the parallel implementation.

\prettyref{fig:strong_scalability_speedup} shows the average speedup
of the parallel implementation for up to 64 worker processes, using
the standard MW method and our MWD approach. The average speedup was
calculated for the three different problem instances, i.e., 1,280,
2,560, and 5,120 transmitters deployed over the target area. The number
of transmitters used in these problem sizes is comparable to several
real-world radio networks that were already deployed in England, e.g.,
Hampshire County with 227 base stations, West Midlands with 414 base
stations, and Greater London Authority with 1,086 base stations \cite{Number_of_base_stations_in_England}.
Note that it is common for a single base station to host multiple
transmitters. 

The plotted average speedup clearly shows the minimal overhead of
the MWD approach when using a small number of worker processes. This
overhead accounts for the final aggregation of the intermediate results
at the DB, which in the MW configuration is performed along worker
processing. Like before, the DB component allows the parallel implementation
to fully exploit the available computing resources when deploying
a larger number of worker processes, until the network-speed limit
is met. Of course, these results are directly correlated with the
wall-clock times shown in \prettyref{fig:strong_scalability_time}.

Another measure to study how well PRATO utilizes the available computing
resources considers the parallel efficiency of the implementation,
i.e., how well the parallel implementation makes use of the available
processor cores. The definition of parallel efficiency is as follows

\begin{equation}
E(NP)=\frac{S(NP)}{NP},
\end{equation}

\noindent where $S(NP)$ is the speedup as defined in Equation (\ref{eq:speedup}),
and $NP$ is the number of cores executing worker processes. \prettyref{fig:strong_scalability_efficiency}
shows the average parallel efficiency of the parallel implementation
for different problem sizes as we increase the number of processing
cores. Like for the speedup measure, we have calculated the average
parallel efficiency from the three problem instances analyzed.

The ideal case for a parallel application would be to utilize all
the available computing resources, in which case the parallel efficiency
would always be equal to one as we increase the core count. From the
plot in \prettyref{fig:strong_scalability_efficiency}, we can see
that the efficiency of the MWD approach is better than in the MW case
for larger number of processes and as long as there is still capacity
at the LAN level. In accordance to the previous analysis, the under
utilization of the computing resources is more significant when the
master process is overloaded (in the MW case) than when the network
infrastructure is saturated (in the MWD case). The lower efficiency
is directly proportional to the number of idle worker processes that
are waiting for the master process (MW case) or for network access
(MWD case).

Overall, the experimental results confirm that the objective of fully
exploiting the available hardware resources is accomplished when applying
our MWD approach, thus improving the scalability and efficiency of
PRATO when compared with a traditional MW method.

\section{Conclusion \label{sec:Conclusion}}

We have presented PRATO, a parallel radio-coverage prediction tool
for radio networks. The tool, as well as the patterns for exploiting
the computing power of a group of networked computers, i.e., a computer
cluster, are intended to be used for spatial analysis and decision
support. The introduced MWD technique, which combines the use of a
database system with a work-pool approach, delivers improved performance
when compared with a traditional MW setup. Moreover, the presented
system provides parallel and asynchronous computation, that is completely
independent of the GIS used, in this case the GRASS environment. Consequently,
a GIS installation is needed on only one of the nodes, thus simplifying
the required system setup and greatly enhancing the applicability
of this methodology in different environments.

The extensive simulations, performed on the DEGIMA cluster of the
Nagasaki Advanced Computing Center, were analyzed to determine the
level of scalability of the implementation, as well as the impact
of the presented methods for parallel-algorithm design aimed at spatial-data
processing. The conducted analyses show that when using the MWD approach,
PRATO is able to calculate the radio-coverage prediction of real-world
mobile networks in a feasible amount of time, which is not possible
for a serial implementation. Moreover, the experimental results show
PRATO has a better scalability than the standard MW approach, since
it is able to completely saturate the network infrastructure of the
cluster. These promising results also show the great potential of
our MWD approach for parallelizing different time-consuming spatial
problems, where databases form an intrinsic part of almost all GIS.
Furthermore, the automatic optimization of radio networks, where millions
of radio-propagation predictions take part in the evaluation step
of the optimization process, are also excellent candidates for this
approach. Indeed, this last point is currently undergoing extensive
research and it is already giving its first results.

Encouraged by the favorable results, further work will include abstracting
the introduced MWD principle into a multi-purpose parallel framework
such as Charm++~\cite{Kale-The_Charm_Approach:2013}, which provides
a functionality for overlapping execution and communication, as well
as fault tolerance.

In addition, as PRATO is also a free and open-source software project%
\footnote{The source code is available for download from http://cs.ijs.si/benedicic/%
}, it can be readily modified and extended to support, for example,
other propagation models and post-processing algorithms. This characteristic
provides it with a clear advantage when compared to commercial and
closed-source tools.

\section*{Acknowledgments}

This project was co-financed by the European Union, through the European
Social Fund. Hamada acknowledges support from the Japan Society for
the Promotion of Science (JSPS) through its Funding Program for World-leading
Innovative R\&D on Science and Technology (First Program).

\end{document}